\documentclass[aps,prl,twocolumn,superscriptaddress,
nobibnotes,nodoi,noeprint,longbibliography]{revtex4-1}
\usepackage{graphicx}
\usepackage{dcolumn}
\usepackage{bm}
\usepackage{float}
\usepackage{color}

\usepackage{refstyle}
\usepackage{mathrsfs}
\usepackage{amsmath}

\usepackage{esint}
\usepackage{ulem}
\usepackage{siunitx}

\usepackage[unicode=true,pdfusetitle,bookmarks=true,
bookmarksnumbered=false,bookmarksopen=false,breaklinks=false,
pdfborder={0 0 0},backref=false,colorlinks=true,citecolor=red]
{hyperref}
\begin{document}
\title{Large scale zigzag pattern emerging from circulating 
active shakers}
\author{Gaspard Junot}
\affiliation{Departament de F\'{i}sica de la Mat\`{e}ria Condensada, Universitat de Barcelona, 08028 Spain}
\author{Marco De Corato}
\affiliation{Aragon Institute of Engineering Research (I3A), University of Zaragoza, Zaragoza, Spain}
\author{Pietro Tierno}
\email{ptierno@ub.edu}
\affiliation{Departament de F\'{i}sica de la Mat\`{e}ria Condensada, Universitat de Barcelona, 08028 Spain}
\affiliation{Universitat de Barcelona Institute of Complex Systems (UBICS), Universitat de Barcelona, Barcelona, Spain}
\affiliation{Institut de Nanoci\`{e}ncia i Nanotecnologia, Universitat de Barcelona, Barcelona, Spain}
\date{\today}
\begin{abstract}
We report the emergence of large zigzag bands in a population of reversibly actuated magnetic rotors that behave as active shakers, namely squirmers that shake the fluid around them without moving. The shakers collectively organize into dynamic structures displaying self-similar growth, and generate topological defects in form of cusps that connect vortices of rolling particles with alternating chirality. 
By combining experimental analysis with particle-based simulation, we show that the special flow field created by the shakers is the only ingredient needed to reproduce the observed spatiotemporal pattern. 
We unveil a self-organization scenario in a collection of driven particles in an viscoelastic medium emerging from the 
reduced particle degrees of freedom, as here the frozen 
orientational motion of the shakers. 
\end{abstract}
\maketitle
Viscoelasticity,
namely the tendency of a material 
to display both viscous and elastic response under external deformation,
is commonly observed in 
a broad range of systems, from
anelastic solids~\cite{Pipkin1964,Jaglinski2007}, to liquid crystals~\cite{Gennes2001}, micelles~\cite{Chen2010}, concentrated  colloidal suspensions~\cite{Mason1995,Hunter2012} biopolymers~\cite{Storm2005,Koenderink2009} or living cells~\cite{Trepat2007,Hang2022}.
In viscoelastic fluids, the internal molecular rearrangement span several time and length scales,
provoking a series of
intriguing phenomena including stress relaxation, hysteresis, 
memory, creep or shear thickening~\cite{Oswald2001}. 
While the bulk behavior of such fluids 
has been the matter of much research to date, 
emergent directions point towards investigating how such 
fluids mediate the organization of dispersed microscopic particles. 
Active~\cite{Gomez2016,Gaojin2016,Chih-kuan2017,Lozano2019,Gaojin2021}, passive~\cite{Smalyukh2005,Kotar2006,Park2016} or externally driven~\cite{Navarro2014,Avino2015,puente2019viscoelastic,Junot2022,Rogowski2021,su_castillo_2022} particles in viscoelastic fluids 
are excellent model systems for many-body organization in elastic materials,
while displaying promising applications in microrheology~\cite{Winter2012,Khan2019}, tissue engineering~\cite{Lee2001,Kwee2017,Cianchetti2018} or microrobotics~\cite{Fusco2014,Palagi2018}.
Indeed fluid elasticity can directly affect 
the propulsion behavior of microswimmers~\cite{Fu2007,Teran2010,Shen2011,Pak2012,Spagnolie2013,Thomases2014,Binagia2021,Arratia2022},
or even be used 
to obtain net motion via streaming flow
when pair of particles interact~\cite{Pak2012,Datt2018,Kroo2022}. 
However, most of these works 
have been focused on single or few interacting particles, leaving the rich physics of ensembles
a rich ground for exploration.

\begin{figure}[th]
\begin{center}
\includegraphics[width=\columnwidth,keepaspectratio]{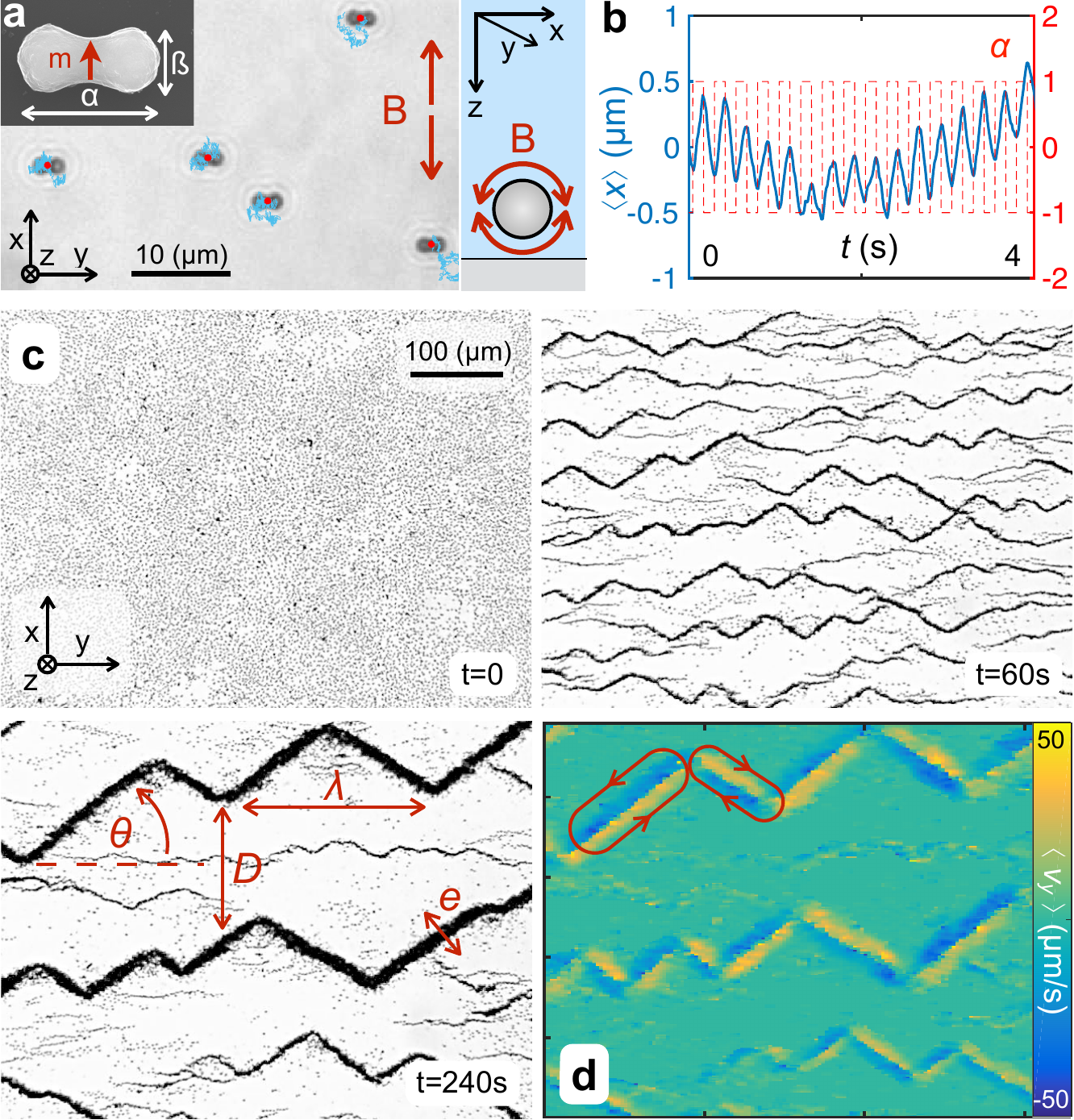}
\caption{(a) Trajectories of a dilute suspension of shakers
after $N=68$ cycles of the rotating field, MovieS1 in~\cite{EPAPS}.
Top inset shows scanning electron microscope image of one hematite particle with the permanent magnetic moment $\bm{m}$. Small scheme on the right side shows lateral view of one magnetic roller. (b) Mean displacement $\langle x \rangle$ versus time $t$
of shakers. 
Dashed line denotes the field direction of rotation $\alpha$. (c) Sequence of images illustrating the band formation from an initially disordered suspension ($t=0$) and under a magnetic modulation with 
$f=80$Hz and $\delta t=1/8$s, see also MovieS3 in~\cite{EPAPS}.
The image at $t=240$ s shows two bands at a distance $D$ 
with wavelength $\lambda$,
bond angle $\theta$ and thickness $e$. (d) Average horizontal velocity $\langle v_y \rangle$ 
of the shakers in bands. Red arrows denote the direction of the circulating particles, MovieS4 in~\cite{EPAPS}.}
\label{figure1}
\end{center}
\end{figure}

Here we demonstrate that 
a collection of active shakers made of driven magnetic microrotors can self organize into large scale dynamic bands displaying a zigzag shape. 
First, we experimentally determine the flow field around a microrotor which originates from a microscopic version of the Weissenberg effect~\cite{Weissenberg1947,Janes1967,Lodge1988}, and is astonishingly similar to that of a shaker force dipole~\cite{Hatwalne2004}. 
Finite element simulations confirm that the elastic stresses around the periodically rotating particles drive a net dipolar flow field.
We then characterize the growth process which starts at a microscopic level with pairs of rotors and grows beyond the millimeter scale. We found that the growth is scale invariant and linear with time. Using particle-based simulations based on a minimal model, we show that the shakerlike flow is at the origin of the instability, and it explains the observed constant angle of the bands. These results
suggest that the formation of zigzag patterns is a general effect which could arise in a broad range of systems. 

We disperse anisotropic microparticles in a solution of 
polyacrylamide (PAAM) at a concentration of 
$0.05\%$ by volume in deionized water; see Supplemental Material (SM) for more details~\cite{EPAPS} which includes Ref.~\cite{kulicke1982preparation,brooks1982streamline,keunings1986high,Massana-Cid2019}. 
The PAAM is a linear, high-molecular weight polymer
[$M_w = 5-6 \cdot 10^6$]
and its addition to water made the solution viscoelastic, 
see SM~\cite{EPAPS} for further details. From previous works \cite{zell2010there,del2015rheometry} we estimate the stress relaxation time of such diluted PAAM solution to be of the order $\tau \sim 3$ms. 
Within our PAAM solution we then disperse 
homemade ferromagnetic hematite colloids, prepared using a sol-gel technique~\cite{Sugimoto1993,EPAPS} and characterized by a peanut-like shape with two lobes with a long (short) axis equal to $\alpha=2.6 \, {\rm \mu m}$ ($\beta=1.2\,  {\rm \mu m}$), see inset in Fig.~\ref{figure1}a.
The particles display a permanent magnetic moment 
of amplitude $m \simeq 9 \cdot 10^{-16} \, \rm{Am^2}$~\cite{Fernando2018} and
oriented perpendicular to their long axis. 
Once dispersed in the PAAM solution, the particles sediment  
due to density mismatch, and float at an almost fixed elevation $h$
due to the balance between gravity and electrostatic repulsion with the close substrate. 

We realize active colloidal shakers by cyclically driving our particles 
back and forward along a fixed direction (here the $\bm{\hat{x}}$ axis) using 
a time dependent rotating field,
\begin{equation}  
\bm{B} = B\left[\sin{(2\pi t \Delta f_{-}})\bm{\hat{x}} + \cos{ (2\pi t\Delta  f_{+})}\bm{\hat{z}}\right] \, \, \, ,
\label{applied_field}  
\end{equation}  
with $\Delta f_{\pm}= f \pm \delta f/2$
and $\delta f$ the frequency difference between the two field components along the $\bm{\hat{x}}$ and $\bm{\hat{z}}$ axis.
The applied modulation periodically changes the direction of rotation $\alpha$ every  
$\delta t = 1/ (2\delta f)$, inducing a magnetic torque
$\bm{\tau}_m=\bm{m} \times \bm{B}$
that sets the particles into rotational motion around their short axis with an angular speed $ \Omega  = 2\pi f$  for frequencies $f<100$Hz (synchronous regime).
Moreover, since the applied field is circularly polarized along the ($\bm{\hat{x}},\bm{\hat{z}}$) plane, it 
aligns the permanent moments of the particles, ensuring a fixed angular orientation. 
Here we fix $B=5.5$mT and $f=80$Hz.

The presence of a solid surface breaks the spatial symmetry, and produces a rolling transport due to the rotation-translation hydrodynamic coupling~\cite{Happel1973}.
In the limit $\delta f = 0$, the field does not switches, and the hematite particles roll above the substrate 
by acquiring a frequency tunable propulsion speed $v_x \sim  2 \pi \beta f$~\cite{Junot2021},
with $\beta$ the particle short axis. 
In contrast, for a finite frequency delay $\delta f =4$Hz, individual particles perform small oscillations of amplitude $\Delta x \sim 0.5 \rm{\mu m}$ and zero average velocity, Fig.~\ref{figure1}(b) herein and Movie S1 in Ref~\cite{EPAPS}.

The small amplitude time-reversible motion of the particle should produce no net displacement in a Newtonian fluid~\cite{Pur97}, 
as shown in Movie S2 in Ref~\cite{EPAPS}. Instead, we find that, at high particle density $\phi=0.117\pm0.002$, the shakers organize in complex dynamic bands which grow linearly with time.
As shown in the sequence of images in
Fig.~\ref{figure1}c, a system of randomly distributed particles evolves into a structured array of bands with zigzaglike shape after few second of magnetic driving.
The bands grow first by
acquiring nearest particles, located on their lateral sides, and then via a continuous merging; see Movie S3 in Ref~\cite{EPAPS}.
These dynamic bands acquire a zigzag shape with branches arranged at a constant angle of $\theta_l = \pm31^{\circ}$ delimited by cusps.
A careful inspection of the particle velocity within a band, Fig.~\ref{figure1}(d), reveals that the shakers move collectively forming  
rotating vortices with fast circulating edge currents up to $50 \rm{\mu m \, s^{-1}}$; see Movie S3 in Ref~\cite{EPAPS}. 
Each branch of a band 
is made of a large scale vortical flow of particles,
and cusps within a band connects vortices of opposite chirality, similar to a two gears system. These bands were observed to form also for 
smaller peanuts ($\alpha = 1.8 \, \rm{\mu m} $, $\beta = 1.3 \, \rm{\mu m} $), or by varying $f \in [40,100]$Hz and for $\delta f >0.75$Hz.

\begin{figure}[t!]
\begin{center}
\includegraphics[width=0.95\columnwidth,keepaspectratio]{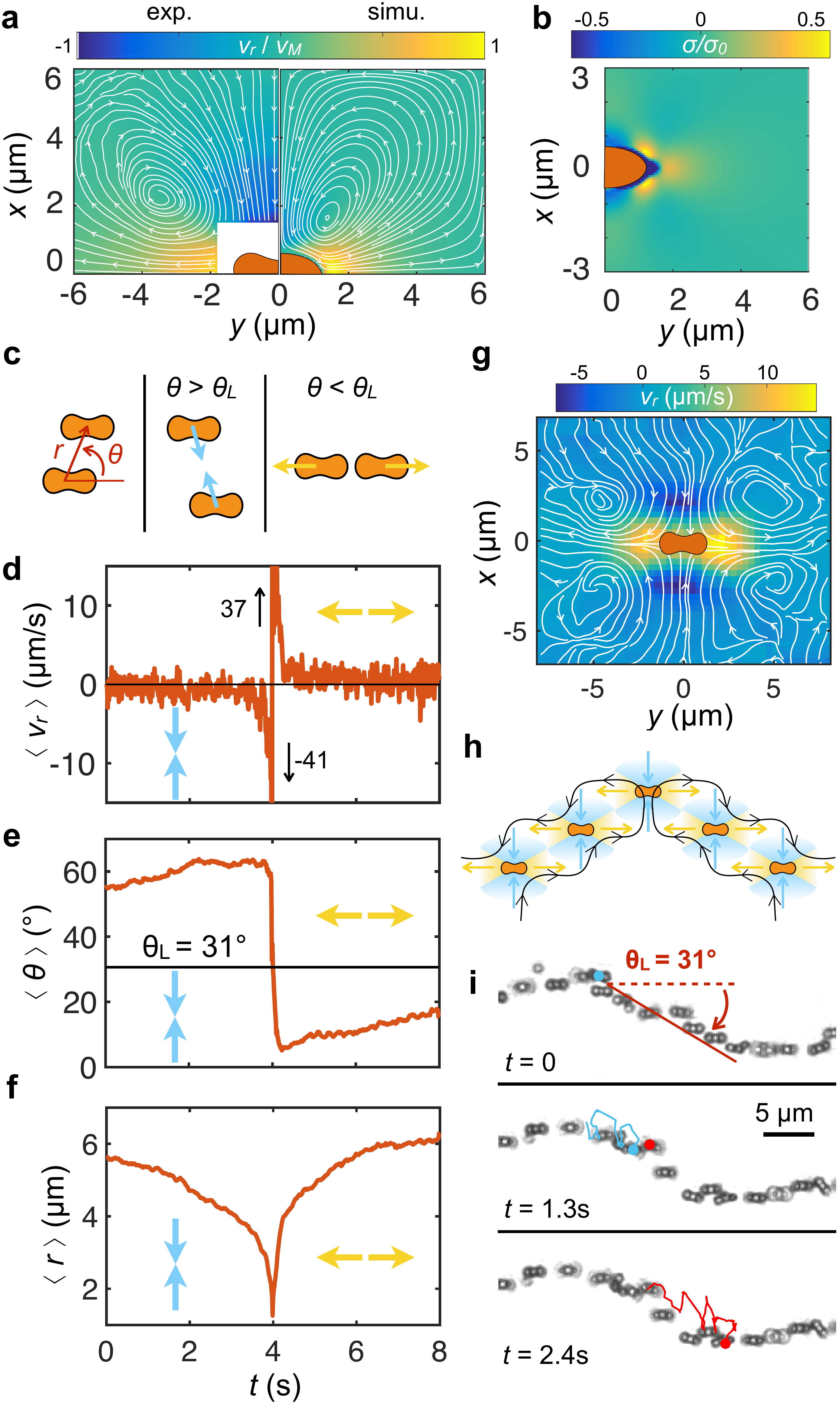}
\caption{(a) Normalized flow field created by a shaker 
$v_r/v_M$  
in the $(x,y)$ plane.
The color codes for the radial velocity 
$v_r$ is normalized by the maximal radial velocity $v_M$. It shows
regions of attractions ($v_r<0$)
and repulsion  ($v_r>0$) separated
by $\theta_l$.
Left (right) panel refers 
to experimentally measured (simulated) 
flow field ($v_M=2.2 \, \rm{\mu m \, s^{-1}}$ for the experiments, $v_M=22\rm{\mu m \, s^{-1}}$ for the simulation see also the SM~\cite{EPAPS}). 
(b) Normalized first normal stress difference in the $(x,y)$ plane with $\sigma_0 = (\eta_s + \eta_p)2\pi f$, see text.
(c) Schematics of two microrotors in three configurations: 
attraction (repulsion) arises when $\theta > \theta_l $
($\theta < \theta_l$) with $\theta_l = 31^{\circ}$.
(d-f) Evolution with time of (d) the average radial velocity $\langle v_r \rangle$, 
(e) the relative angle $\langle \theta \rangle$
and (f) the radial distance $\langle r \rangle$
between two approaching microrotors.
In all images blue [black] (yellow [gray]) arrows indicate
attraction (repulsion) 
between the pair, MovieS5 in~\cite{EPAPS}. 
(g) Top view of the flow velocity generated from the interaction of two shakers. 
(h) Assembled structure from the flow produced by the shakers. 
(i) Sequence of snapshots showing a one-particle thin
band with superimposed two particle's trajectories, MovieS6 in~\cite{EPAPS}.}
\label{figure2}
\end{center}
\end{figure}

To understand these unexpected, complex structures we analyzed the flow field generated by a single microrotor. We obtain the velocity field around a microrotor, averaged over many field periods, via particle tracking velocimetry; see SM for details~\cite{EPAPS}.
Note that in a Newtonian fluid hydrodynamics interactions are time reversible so that no net particle displacement would emerge. 
Strikingly, the obtained flow field in Fig.~\ref{figure2}(a), 
displays strong similarities with that generated by a shaker-like force dipole \cite{Hatwalne2004}. 
Shakers are a category of squimers that generate a flow pattern similar to that of several microorganisms such as \textit{Escherichia coli} bacteria \cite{Lauga2009} but without the polar component, which prevent them from self-propelling. These particles exert stresses on the surrounding fluid
\cite{hatwalne2004rheology,chaithanya2020deformation,scagliarini2022hydrodynamic}
that drive a flow field displaying an attracting part at the particle sides and a repulsive one at the tips, with a recirculation vortex between these two regions. 
An estimate of the Deborah number, $De = 2 \pi \, f \, \tau $,  yields $De \approx 1.5$, which suggests that this particular flow field results from the fluid elasticity. Thus, the particle rotation induces a normal stress difference along the $\bm{\hat{x}}$-axis which induces a flow toward the rotor and, by volume conservation, the fluid is expelled toward the tips; Fig.~\ref{figure2}(b). 
We confirm this hypothesis by computing the flow velocity $\mathbf{v}$ around a periodically-rotating ellipsoid via three-dimensional numerical simulations.  We consider a stress tensor 
$\mathbf{T}=-p\mathbf{I}+\eta_s\left(\mathbf{\nabla}\mathbf{v}+{\mathbf{\nabla}\mathbf{v}}^T\right)+\mathbf{\sigma} \, \, 
$ as sum of a Newtonian contribution (viscosity of water $\eta_s$) coming from the solvent and a viscoelastic one $\mathbf{\sigma}$ introduced by the PAAM. Here $p$ is the pressure that enforce the incompressibility condition,
$\mathbf{\nabla} \cdot \mathbf{T}=\mathbf{0}$. We use 
the Oldroyd-B constitutive model, which predicts a constant viscosity but a non zero normal stress difference~\cite{larson2013constitutive}, via an additional 
constitutive equation:
\begin{equation} 
\tau\left(\frac{\partial}{\partial t}\mathbf{\sigma}+\mathbf{v}\cdot\mathbf{\nabla}\mathbf{\sigma}-\mathbf{\nabla}\mathbf{v}\cdot \mathbf{\sigma}-\mathbf{\sigma}\cdot{\mathbf{\nabla}\mathbf{v}}^T\right)+\mathbf{\sigma}=\eta_p\left(\mathbf{\nabla}\mathbf{v}+{\mathbf{\nabla}\mathbf{v}}^T\right) \, \, ,
\end{equation}
where $\eta_p$ is the polymer viscosity.
Such model reduces to only two parameters, $\eta_p$ and $\tau \sim 3$ ms. The polymer viscosity can be extracted from the zero-shear viscosity of the solution, which from the experimental measurements is $\eta_0\approx{2\ \eta}_s$; thus, we estimate $\eta_p\approx\eta_s$. See SM~\cite{EPAPS}
for further technical details.

The right panel of Fig.~\ref{figure2}(a) shows that the computed velocity field in the plane $z=h$ and averaged over one period $\delta t$ displays the same features as that measured in the experiments and confirms that the mechanism driving the dipolar flow structure is the first normal stress difference; see 
Fig.~\ref{figure2}(b). This result is consistent with the seminal works by Giesekus \cite{giesekus1965some} and Fosdick and Kao \cite{fosdick1980steady}, who showed that the first normal stress distribution around a rotating sphere generates streamlines that increase with even powers of $De$. It follows that the flow field introduced by the first normal stresses does not change sign upon changing the direction of the particle rotation leading to a non zero average over one period.

\begin{figure}[b]
\begin{center}
\includegraphics[width=\columnwidth,keepaspectratio]{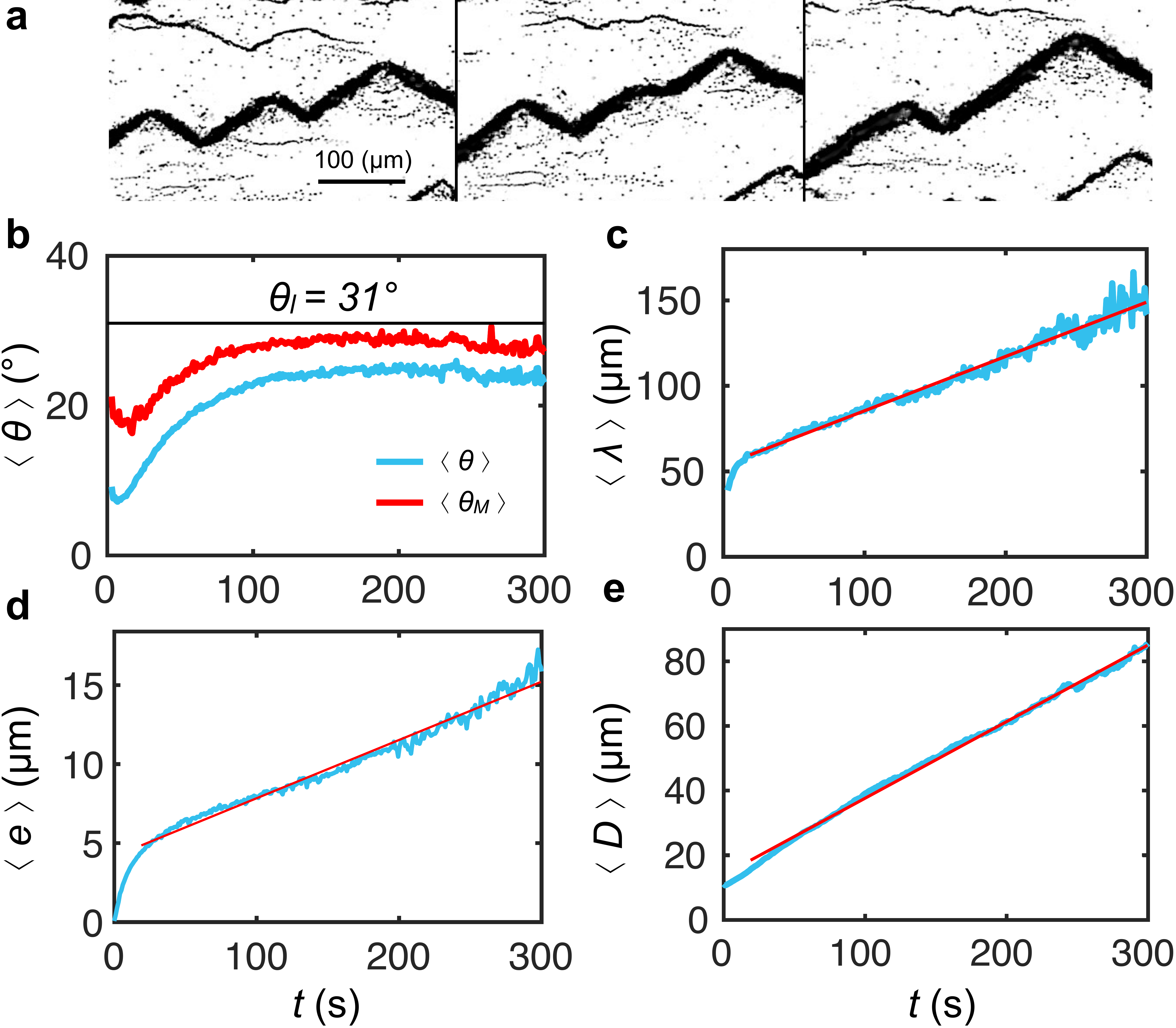} 
\caption{
(a) Sequence of snapshots showing the annihilation of a cusp (total duration $\Delta t=30$ s), MovieS7 in~\cite{EPAPS}.
Coarsening dynamics: time evolution of the mean angle $\langle \theta \rangle$ (b),
wavelength $\langle \lambda \rangle$ (c),
thickness $\langle e \rangle$ (d) 
and distance $\langle D \rangle$ (e).
After a short transitory, $\lambda$, $e$ and $D$
all scale linearly with time (red line) while $\theta$ set to a stationary value of $\theta_l=31^\circ$.
In (b) $\langle \theta \rangle$ corresponds to the full average over the experimental data, weighted by the size of the zigzag bands, while  $\langle \theta_M \rangle$ is its maximal value over the bands.}
\label{figure3}
\end{center}
\end{figure}

We then analyzed the interactions between a pair of shakers using data from $33$ separate experiments.  
Because of the field alignment, the magnetic particles display negligible orientational motion  
and the relative orientation can be described 
in term of a single angle $\theta$, Fig.~\ref{figure2}(c). 
The relative velocity field between the pair has a similar structure  than the one generated by a single one. Two particles attract each other when are side by side and repel when positioned tip-to-tip. The transition between the two regions occurs at an angle $\theta_l = 31^{\circ}$, Fig.~\ref{figure2}(c). 
When two particles 
\begin{figure}[!t]
\begin{center}
\includegraphics[width=\columnwidth,keepaspectratio]{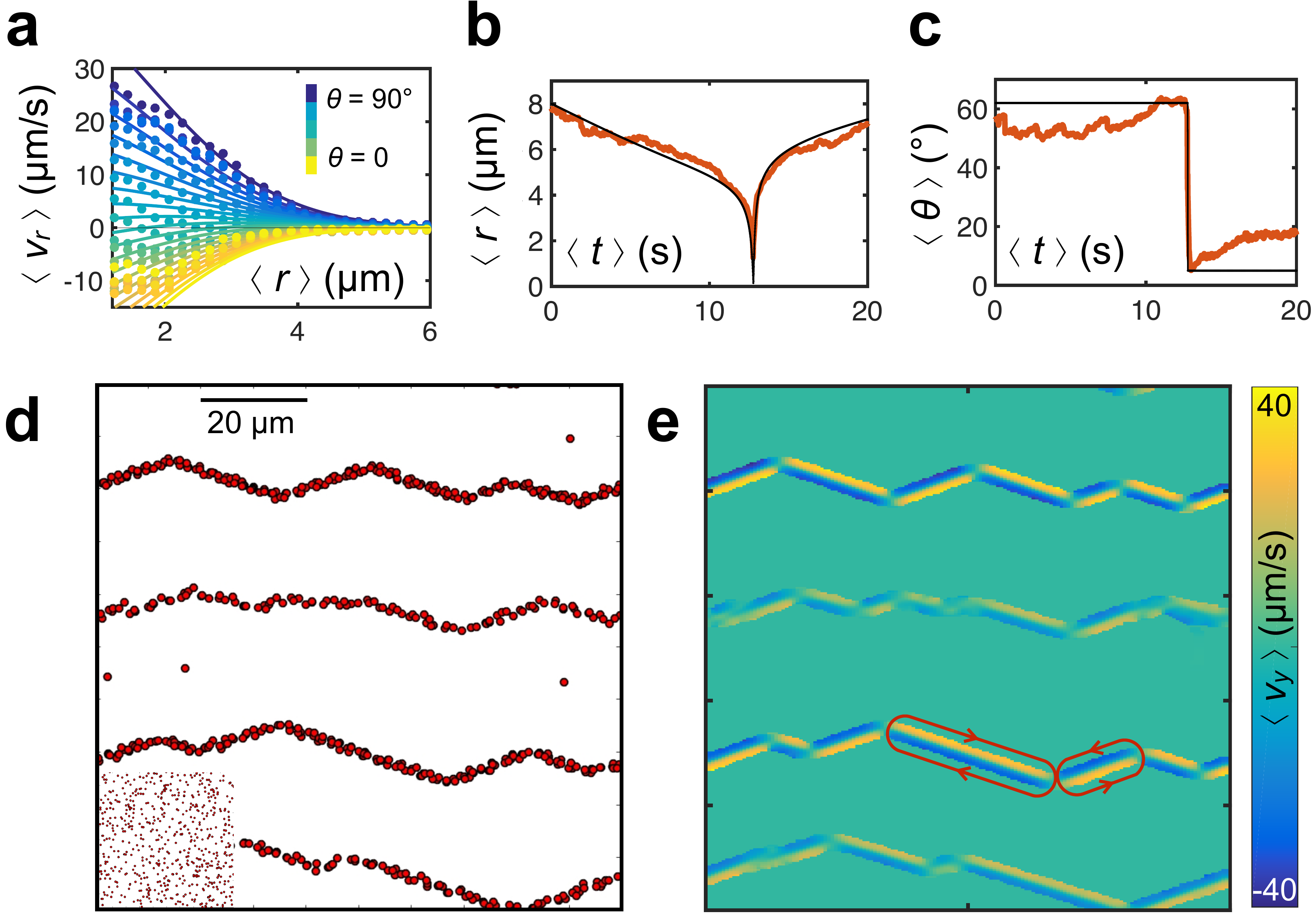} \caption{(a) Average radial velocity $\langle v_r \rangle$ between two particles as function of their relative distance $r$ and for different angle $\theta$ (colorbar). Scattered data are experiments while continuous line are non-linear regression using Eq. 9 in~\cite{EPAPS}. (b,c) Average radial distance (b) and relative angle (c) as function of time for pairs of interacting particles. In both graphs, experiments (simulation) are represented by the red [gray] (black) curve. (d) Image showing the zigzag state obtained after a time $t=50$ s with $N=601$ particles, MovieS8 in~\cite{EPAPS}. Small inset shows the initial random configuration of the particles. (e) Corresponding color-coded plot showing the average horizontal velocity $\langle v_y \rangle$ with  arrows denoting the direction of the circulating particles.}
		\label{figure4}
	\end{center}
\end{figure}
are initially arranged such that $\theta>\theta_l$,
they approach first slowly
and keeping their relative orientation constant.
Near close contact, $r=2 \rm{\mu m}$ we observe  
a rapid sliding process which causes a speed up 
effect reaching relative velocities up to $v_r=40 \rm{\mu m s^{-1}}$,
Figs.~\ref{figure2}(d)-(f). 
Such a process
re-arrange the rotors from 
side-by side ($\theta \sim 60^{\circ}$)
to tip-to-tip ($\theta \sim 0^{\circ}$),
Movie S5 in Ref~\cite{EPAPS}. 
Thus, close particles arrange themselves at an angle $\theta_l$ where attraction and repulsion are minimized and, by drawing the flow lines, one recovers the direction of rotation of the vortices, Fig.~\ref{figure2}(h). During band growth, defects arise in form of cusps which sink incoming particles expelling them from the opposite side.
Such a hypothesis is confirmed by observing the formation of one line thin band, Fig.~\ref{figure2}(i)
and Movie S6 in Ref~\cite{EPAPS},
where the constituent particles detaching from the 
branch are dragged along the vortical edge current toward the nearest cusp. 

The ensemble of shakers exhibit a self-similar behavior with scale invariance in time as they evolve 
to large scale structures, Fig.~\ref{figure1}(c). 
During coarsening, the different band parameters as wavelength [$\lambda$, Fig.~\ref{figure3}(c)], thickness [$e$, Fig.~\ref{figure3}(d)] and distance [$D$, Fig.~\ref{figure3}(e)] grow linearly in time, while the the bond angle rapidly saturate to $\theta_l$, Fig.~\ref{figure3}(b) \footnote{To avoid that the averaged values are dominated by the small structures that are more numerous than the large ones but only contain a small fraction of the particles, 
$\langle \theta \rangle$, $\langle \lambda \rangle$ and $\langle e \rangle$ are weighted by the area of the zigzag band. Thus, small zigzag bands count less than larger ones.}.
As shown in Fig.\ref{figure3}(a), a band is composed of a sequence of topological defects 
in form of cusps that connect different branches in a zigzag manner. 
During coarsening, small branches disappear in favor of large one inducing the annihilation of  cusps, Movie S7 in Ref~\cite{EPAPS}.
This dynamically slow process ultimately would lead to a single straight band of particles, or to separate bands at very large distance between them. However, due to the slow velocity of coarsening and the large system size (observation window $0.66$ mm, whole system size $1$ cm), this state is difficult to reach experimentally. 
During coarsening, separate bands merge by reducing their distance until touching each other. Such a process is triggered in part by the cusp annihilation that increases the wavelength $\lambda$ and so the spatial extension of a band along the lateral direction ($\bm{\hat{x}}$). The fusion of two bands increases the inter-band distance $D$ and give rise to a new structure with larger thickness $e$. This new band will in turn annihilates its cusps, starting a new cycle. 
The cusp's annihilation process ($[A]+[A] \rightarrow 0$) can be written as $\frac{d[A]}{dt}=-k[A]^2$, being $[A]$ the linear cusp concentration in a band and $k$ the annihilation rate constant. The solution of this equation is $[A](t) = a_0/(1+a_0kt)$ with $a_0 = [A](t=0)$. By definition, $\lambda = 2/[A]$, one thus recovers
the linear growth in time of the wavelength by writing,
$\lambda(t)=2(kt+1/a_0)$.

To rationalize our results, we set-up a minimal simulation scheme
that reproduces the self-organization scenario
neglecting magnetic and steric interactions.
The former are not considered given the relative low value of $\bm{m}$. For two hematite rotors $(i,j)$ aligned tip to tip (side by side) at the closest distance of $x_{ij}= \beta$ ($y_{ij}=\alpha$) the time averaged potential is relatively weak, given by $\langle U_{d} \rangle =-\mu_0 m^2/(8 \pi x_{ij}^3)=-5.8 \, k_B T$ ($\langle U_{d} \rangle  =\mu_0 m^2/(4 \pi x_{ij}^3)=1.1 \, k_B T$ resp.), being $\mu_0=4 \pi 10^{-7}\, \rm{H m^{-1}}$. 
Thus, we are left to consider the generated velocity field
[see Fig.~\ref{figure4}(a)] that we obtain directly from the experimental data. 
We use 
an empirical function 
which simultaneously fit all data [Eq. 9 in SM~\cite{EPAPS}]
and provides the relative velocity $\bm{v}(r_{ij};\theta_{ij})$ between particles at relative position $r_{ij}$ and 
orientation $\theta_{ij}$. Thus, we numerically integrate the equation of motion $\frac{d\bm{r}_i}{dt}=\sum_{i\neq j}\bm{v}(r_{ij};\theta_{ij})$, more details on the implementation can be found in the SM~\cite{EPAPS}. Taking into account the different approximation used, the result is rather striking since it allow to reproduced  the band formation process and predict the correct 
circulation flow across the bands, as shown in Figs.~\ref{figure4}(d) and \ref{figure4}(e). Thus, the zigzag pattern arise due to the shakerlike shape of the flow field generated by the particle rotation, Movie S8 in Ref~\cite{EPAPS}.  

In conclusion, we show that zigzag bands emerge because of the shakerlike flow field generated by the magnetic rotors. In our system, this flow field results from the elastic stress created by the particle rotation within the PAAM solution.
However, similar patterns have been also reported in Newtonian fluids for particles submitted to an AC field \cite{jennings1990electro}. The physical origin of this instability was explain by macroscopic gradients in the electrolyte concentration due to the field-induced concentration gradients near the particles surfaces \cite{isambert1997,isambert19972}. Further, such instability was attributed to mutual polarization of particles, causing them to rotate \cite{hu1994observation,lele2008anomalous}, and recently to the presence of electrokinetic flows \cite{katzmeier2021emergence}. 
Our work shows that the zigzag instability is even more general, suggesting that it is not the the forcing (magnetic or electric) nor the medium (Newtonian or viscoelastic) that matter but rather the type of hydrodynamic flow those systems create around the particle.
The frozen orientation of the shakers plays an important role in stabilizing the zigzags. Indeed, in active nematics it was  shown that freezing one orientation via a constant magnetic field induces a zigzag stripe phase~\cite{Guillamat2016}. 
In contrast, systems of force-dipoles free to rotate and/or able to self propelled may not exhibit such stable structures as they generate turbulent dynamics. 
Thus, we find that the
zigzag instability is a general phenomenon that would 
depend only on the symmetry of the flow field developed by
the particles. 
A potential technological application of our work could be to
use the vortices as a conveyor belt to transport particles or to mix fluids at small scales. It could also be used to localize magnetic inclusions in microfluidc devices. The oscillating field could be used for example to control the flow by inducing (removing) clogging when oscillating field is switched on (off).

We thank Thomas M. Fischer and Jaume Casademunt  for stimulating discussions, and Jordi Ort\'in for help with the rheological measurements. This work has received funding from the European Research Council
(ERC Consolidator Grant contract No. 811234). 
P. T. acknowledge support from the program "ICREA Acad\`emia".
M.D.C. acknowledges funding  the Spanish Ministry of Science and Innovation (MCINN) under the Juan de la Cierva (IJC2018-035270-I) postdoctoral fellowship and the retos de investigaci\'on grant PID2020-113033GB-I00.


\section{Supplementary Material}

\subsection{Experimental details}
We prepare the PAAM solution by adding $0.1$g of PAAM to $19.9$ml of milliQ water (MilliQ, Millipore) and mix it via magnetic stirring at $250$rpm during $24$ hours. 
The resulting suspension is further diluted with water until a final PAAM concentration of $0.05\%$. 
At this concentration, the PAAM solution is in the dilute regime (where the polymers do not overlap) close to the transition between dilute and semidilute regime (where the polymers overlap without entanglement). With our experimental conditions, this transition should occurs at a concentration $c^* = 0.06-0.1\%$ \cite{zell2010there,del2015rheometry}.

Due to the very low concentration of PAAM used in experiments, rheological properties could not be measure with a classical rheometer and we characterized more concentrated PAAM solution, see below. 
The solution is then  mixed with a suspension of monodisperse hematite particles characterized by a peanut-like shape.
The particles were previously synthesized using the sol-gel Method~\cite{Sugimoto1993}.
A drop of the resulting suspension is then squeezed between a plastic petridish and a cover slip and later sealed. The experimental cell have a final height of $260 \rm{\mu m}$, and is placed on the stage
of a custom made optical microscope. 
The particle sediment towards the bottom plate due to density mismatch, and the final surface concentration is $\phi= 0.117 \pm 0.002$.
We visualize the system dynamics using an area scan charge-coupled device camera (Scout scA640-74f, Basler) connected to a PC station. The external oscillating magnetic field was applied using using a set of custom-made magnetic coils with their main axis aligned along the three orthogonal directions ($\bm{\hat{x}},\bm{\hat{y}},\bm{\hat{z}}$). A rotating magnetic field in one plane perpendicular to the  substrate was generated by connecting 
two pairs of coils to a power amplifier (IMG STA-800, Stage Line)  commanded by an
arbitrary waveform generator (TGA1244, TTi).

\begin{figure*}[t]
\centering
	\includegraphics[width=0.8\textwidth]{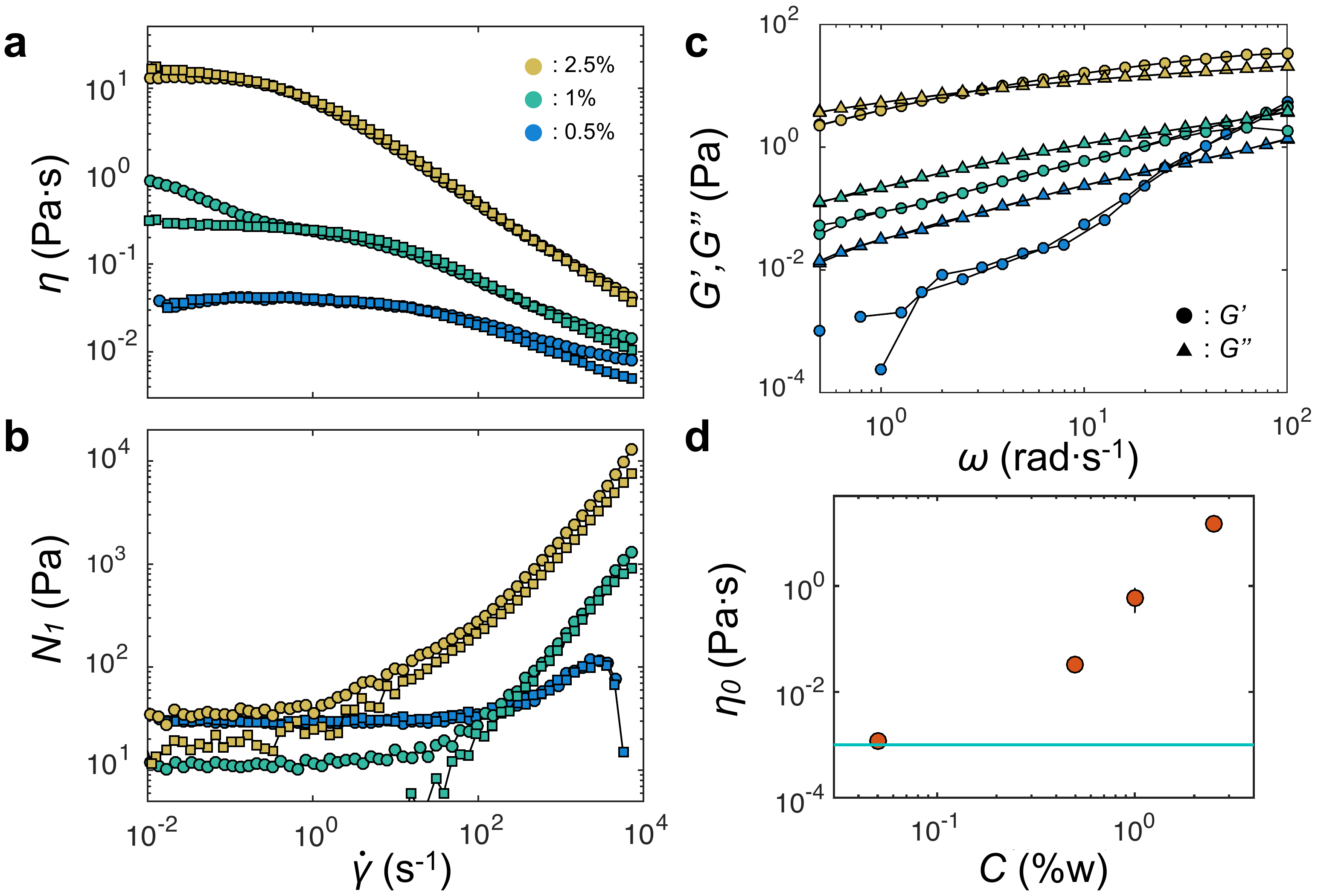}
	\vspace{0 mm}
\caption{Rheology of PAAM solutions of different mass fraction. (a) Viscosity and (b) Normal stress as function of shear rate. Circle (square) symbols represent an increasing (decreasing) ramp of the applied shear rate. (c) Storage (circle) and loss (triangle) modulus as a function of the pulsation. Increasing and decreasing ramps are represented with the same colour and symbol.  (d) Viscosity of water and PAAM solution
measured by considering the attraction of two magnetic 
particles having relative velocity $V$. (e) Zero shear viscosity 
$\eta_0$ of PAAM solutions
as function of the PAAM concentration $C$ 
(percentage in mass).}
\label{Fig_rheology}
\end{figure*}

\begin{figure}[t]
\centering
	\includegraphics[width=0.9\columnwidth]{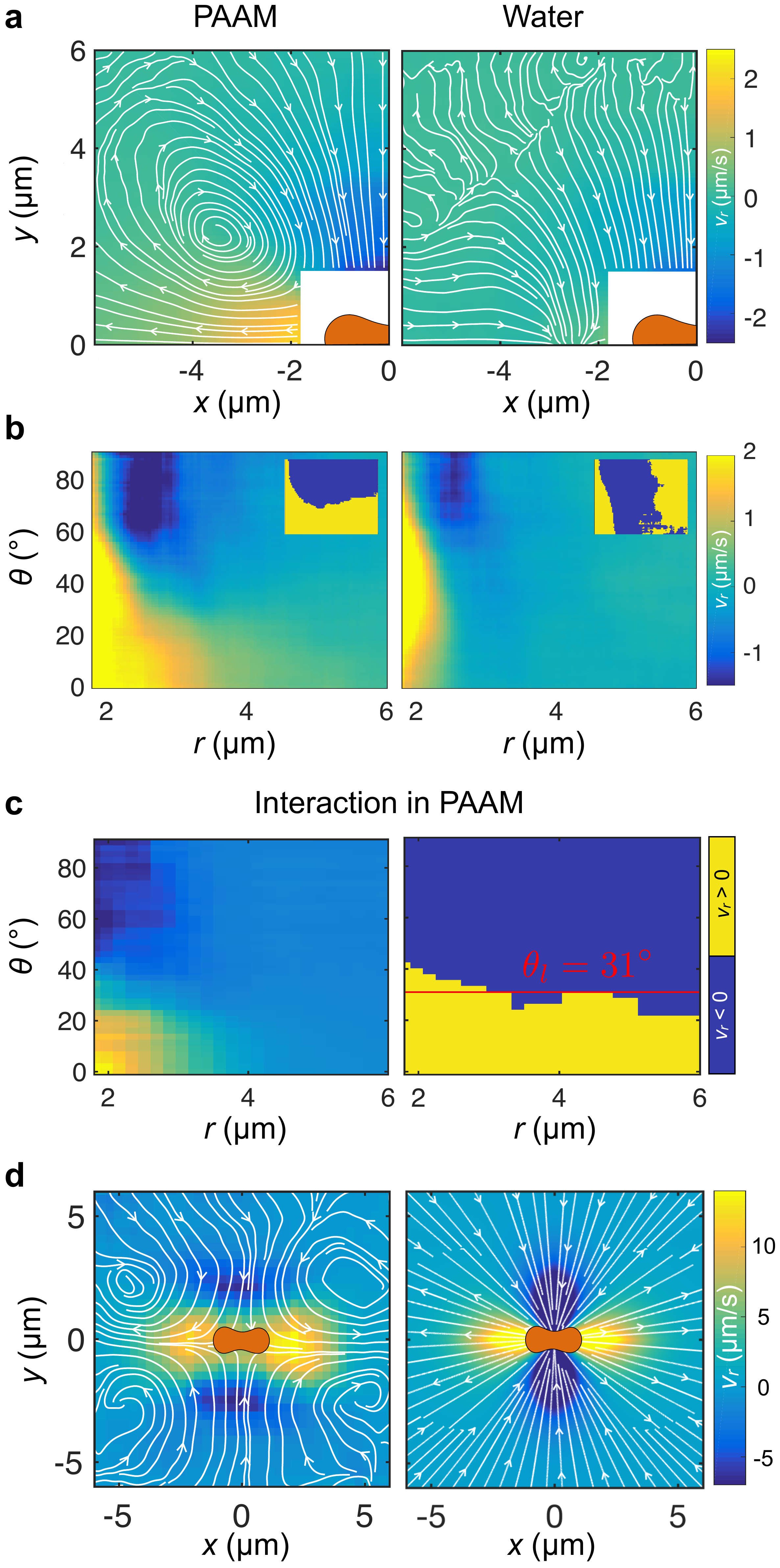}
	\vspace{0 mm}
\caption{(a,b) Flow field around an individual microrotor in the $(\bm{\hat{x}},\bm{\hat{y}})$ plane (a) and in 
the $(\bm{\hat{r}},\bm{\hat{\theta}})$ plane (b).
The left  (right) panels refers to PAAM (water) solution. Insets show the binarized flow field which is colored yellow for $v_r > 0$ and blue for $v_r < 0$. (c) Left: Flow field generated by the interaction of two microrotors in the $(r,\theta)$ plane. Right: Corresponding binarized flow field, the line $\theta_l=31^{\circ}$ splits the attractive region from the repulsive one. 
(d) Flow field generated by the interaction of two microrotors
in the $(x,y)$ plane. Right: same figure as Fig.2(f) in the main text, left: flow
field generated from Eq.(4) in the Method Section. 
For all figures, the color codes indicate the intensity of the radial velocity.}
\label{Fig_SI_interaction}
\end{figure}

\begin{figure}[t]
\centering
	\includegraphics[width=\columnwidth]{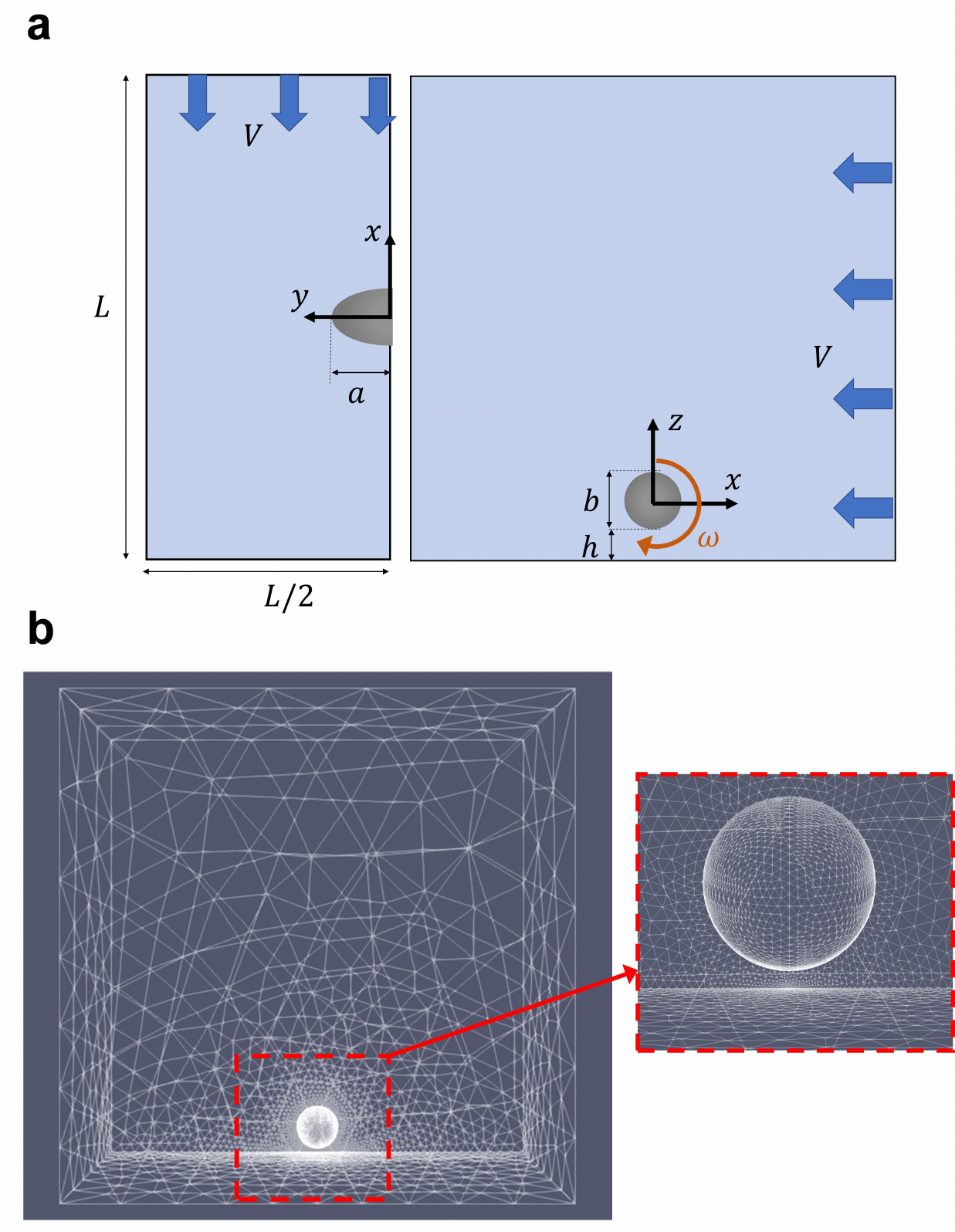}
	\vspace{0 mm}
\caption{Schematics of the computational domain. (a) Schematic drawing of of the geometry considered for the finite element simulations. (b) The tetrahedral mesh used to solve the problem is more refined near the rotating ellipsoid and in the gap between the particle and the bottom wall.}
\label{Fig_Marco_simu_schem}
\end{figure}

\begin{figure}[t]
\centering
	\includegraphics[width=\columnwidth]{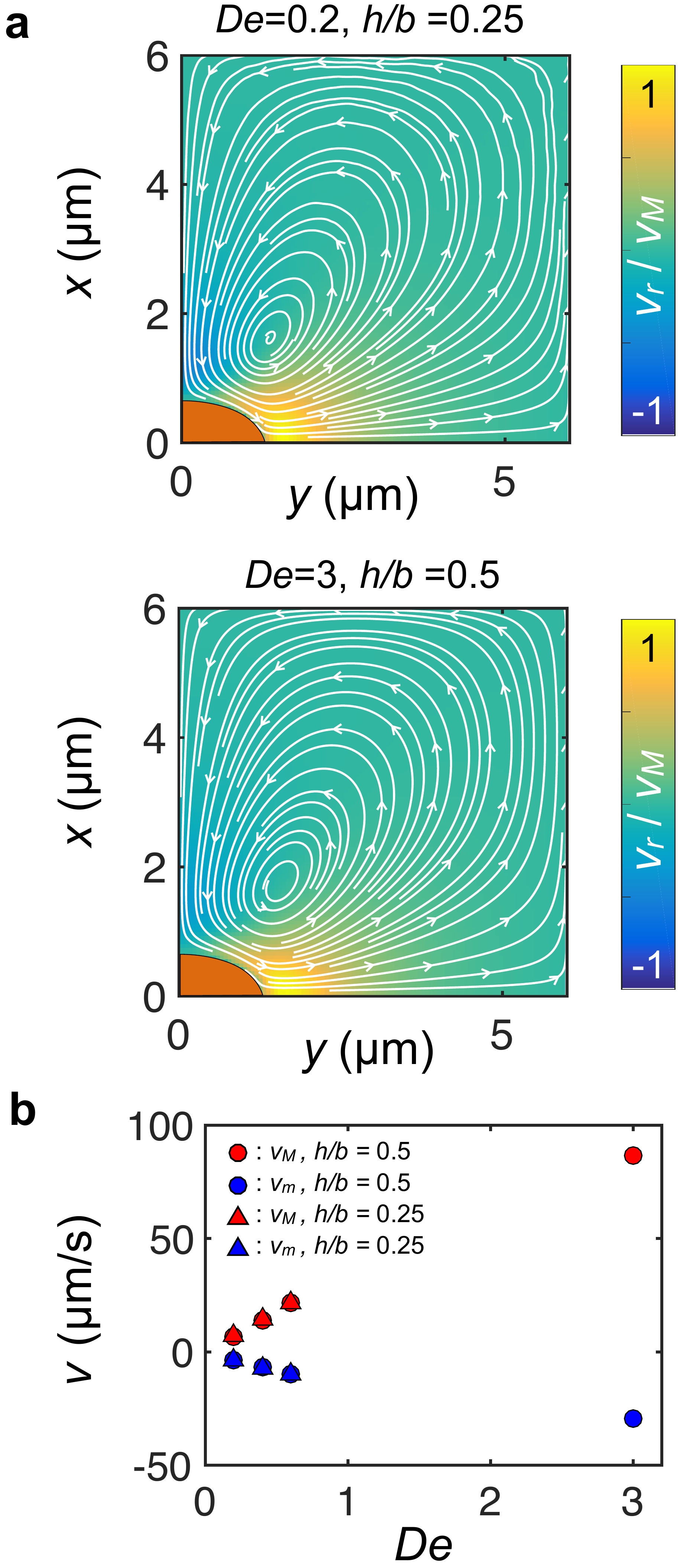}
	\vspace{0 mm}
\caption{Simulation of the viscoelastic flow around a rotating particle. (a) Flow field in the plane $z=0$ for different Deborah number $De$ and normalized particle elevation $h/b$ with respect to the surface (b is the particle minor axis), the colour corresponds to the radial velocity normalized by the maximal velocity. (b) Minimal and maximal radial velocity for different Deborah number and normalized particle elevation.}
\label{Fig_Marco_simu}
\end{figure}

\subsubsection*{PAAM rheology}
We characterized the rheological properties of different PAAM solutions characterized by various PAAM mass fraction. Such properties were then measured using a DHR-2 rheometer (TA Instruments) with a cone-plate geometry. Fig.\ref{Fig_rheology}(a) shows the viscosity of the PAAM solution as function of the applied shear rate, with a shear thinning behaviour emerging at large shear rate. As shown in Fig. \ref{Fig_rheology}(b), these 
solutions also exhibit a normal stress than increases with the shear rate and the PAAM concentration. Dynamical rheological measurements were also performed to obtain  the storage $G'$ and the loss $G''$  modulus as a function of the angular frequency of the rheometer, $\omega$. The PAAM solutions exhibit a viscoelastic behavior where the viscous (elastic) part dominates at small (large) frequency and a crossover frequency between both moduli. Increasing the mass fraction of polymer raises both $G'$ and $G''$ and reduces the frequency where the crossover occurs. The observed behaviour is in agreement with previous results on similar polymer solutions \cite{kulicke1982preparation}.

\subsubsection*{Flow field measurements and analysis}

To compute the velocity of the vortex (Fig. 1(d) of the article), we have tracked individual peanut particles in one of our 19 videos and compute their velocity. The image is then divided in cells of width $d_x = \SI{4.4}{\micro\meter}$ and height $d_y = \SI{1.6}{\micro\meter}$. Inside those cells the local mean velocity along $x$ and $y$ are computed by averaging over all the particles inside the cell and over $500$ frames. We then performed a spatial moving average with a windows of size $3d_x \times 3d_y$. 
To obtain the velocity fields of Fig.2(a) and Fig.2(g) of the article videos have been recorded at $75$fps using an oil immersion $100\times$ Nikon objective.
To compute the flow field of Fig.2(a), we prepared a solution of peanut particle at a low particle concentration so that the peanuts do not interact with each other. We then seeded this solution with silica colloids of $\SI{1}{\micro\meter}$ diameter that we used as tracers. We then took $26$ videos of $1$min and tracked the position of both the peanut and the tracers. We then consider a square region of $\SI{10}{\micro\meter}$ around the peanut. This region is divided in square cells of size $d=\SI{0.05}{\micro\meter}$. Inside those cells the local mean relative velocities (velocity along x and y and radial velocity) between the peanut and the tracers are computed by averaging over all the particle inside the cell, over the 1min of the video and over the $26$ videos. We then performed a spatial moving average with a square windows of size $21d = \SI{1.05}{\micro\meter}$.
To compute the flow field of Fig.2(g), we record videos of pairs of interacting peanut particles. We then tracked the positions of the two particles and compute the relative velocities (velocity along $x$ and $y$ and radial velocity). 
We then consider a square region of $\SI{8}{\micro\meter}$ around one of the peanut. This region is then divided in square cells of size $d=\SI{0.16}{\micro\meter}$ to compute the velocity along the $x$ and $y$ directions, and $d=\SI{0.2}{\micro\meter}$ to compute the radial velocity. Inside those cells the local mean relative velocities (velocity along $x$ and $y$ and radial velocity) between the two peanuts are computed by doing a time average over all the video duration and averaging over $33$ different pairs of peanuts. We then performed a spatial moving average with a square windows of size $15d$.

To characterize the growth process (Figures 3(b-e) in the manuscript), we performed $19$ independent experiments, starting each time with a new sample and an homogeneous initial distribution of particles. We then recorded videos at $75$fps with a $10\times$ Nikon objective from which we extracted the different band parameters that we averaged over all bands at a given time and over all the videos.

\subsubsection*{Finite element simulations}
In this section, we report the details on the finite element simulations used to compute the flow field around the rotating particles.
We consider a single prolate ellipsoid with minor axis $b=1 \, \rm{\mu m} $ and major axis $a=2 \, \rm{\mu m}$ that lies parallel to a rigid wall and spins along its major axis at an angular frequency of $\omega=502.65 \, \rm{rad} \, \rm{s^{-1}}$. We use a Cartesian reference frame with origin in the particle centre and that translates with the particle (See  Fig.\ref{Fig_Marco_simu_schem}), where we have used the fact that the  the plane $y=0$ is a plane of symmetry. We assume that the particle-wall distance, $h$, is fixed by the balance of the gravity force, hydrodynamic forces and particle-wall repulsive interactions as $h=b/4$, which corresponds to $h\approx250 \, \rm{nm}$. As a result of the hydrodynamic interactions with the bottom wall, the particle translates in the $x$ direction at a velocity $V$.

The governing equations in the main text are solved in the computational domain shown in Fig.\ref{Fig_Marco_simu_schem}, where we exploited the symmetry plane of the problem $y=0$. In the co-moving frame, the particle position does not change and the fluid far from the particle moves at a velocity $-V$ in the x direction. Note that $V$ is itself an unknown of the problem and it is computed together with the velocity field, the pressure field and the viscoelastic stresses. This choice circumvents the need of deforming the mesh and greatly simplifies the simulations. The rigid wall is positioned at $z=-h$ and lateral size of the domain $L$ is chosen sufficiently large to avoid finite-size effects $L=20 \, b$. 

The governing equations are supplemented with the force balance on the rotating particle:
\begin{equation} 
\int_{S}\left(\mathbf{T}\cdot\mathbf{n}\right){\cdot \bm{\hat{x}}} \, dS=0  \, \, ,
\end{equation}
where we denoted with $S$ the surface of the particle, with $\mathbf{n}$ the vector normal to the surface and pointing inside the fluid domain and with $\bm{\hat{x}}$ the unit vector in the $x$ direction.

The boundary conditions on the bottom wall $z=-h$ is given by the no slip condition
\begin{equation} 
\mathbf{v}=-V \bm{\hat{x}} ,
\end{equation}

On the particle surface the velocity is given by the rigid body rotation generated by the magnetic field
\begin{equation} 
\mathbf{v}=\omega\ \left(z \ \bm{\hat{x}}-x\ \bm{\hat{z}}\right) \, \, ,
\end{equation}
where $\bm{\hat{z}}$ the unit vector in the $z$ direction. 
Neumann boundary conditions are applied to the symmetry plane $y=0$
\begin{equation} 
\mathbf{T}\cdot \mathbf{n}=\mathbf{0} \, \, .
\end{equation}
Finally, we assume no slip boundary condition on the remaining walls
\begin{equation} 
\mathbf{v}\cdot\mathbf{n}=-V \bm{\hat{x}}\cdot\mathbf{n} \, \, .
\end{equation}
The specific boundary condition used on the walls far from the particle plays only a minor role because they are placed far from the rotating particle, $L \gg a $. 

In the experiments the particles are driven by periodically by the magnetic field at at angular velocities $\omega = \pm 502.65 \, \rm{rad} \, \rm{s^{-1}}$, with a period  $T\approx 0.25$ s. Since the period is much longer than the relaxation time, $T \gg \tau$, the viscoelastic stresses adjust very quickly to the change of rotation direction and maintain their steady-state value for most of the period. This greatly facilitates the solution of the equations because we can ignore the transient build up of viscoelastic stresses during the reverse of the field and we directly solve the governing equations assuming steady-state. 
In the experiments, the trajectories of the tracers around the rotating particle are obtained by averaging over many periods in which the particle has changed rotation rate. These trajectories can be obtained from the simulated velocity fields by averaging the results obtained for   $\omega =502.65 \, \rm{rad} \, \rm{s^{-1}}$  and for $\omega =-502.65 \, \rm{rad} \, \rm{s^{-1}}$. Since changing the sign of the rotation rate is equivalent to reversing the direction of motion from $\bm{\hat{x}}$ to $-\bm{\hat{x}}$, we obtain the average over a period simply by
\begin{equation} 
\langle v_x(x,y,z) \rangle =v_x(x,y,z)-v_x(-x,y,z) \, \, ,
\end{equation}  
and
\begin{equation} 
\langle v_y(x,y,z) \rangle=v_y(x,y,z)+v_y(-x,y,z)  \, \, ,
\end{equation}
where the brackets $\langle \rangle$ denote a quantity averaged over one period.

We divide the computational domain in tetrahedral elements with a more refined mesh around the rotating particle. We use a more refined grid in the thin gap between the particle and the wall (See Fig.\ref{Fig_Marco_simu_schem}). We then use the finite element method to discretize the equations and we use an upwind stabilization scheme to stabilize the advective term in the constitutive equation \cite{brooks1982streamline}. We use a quadratic shape function for the velocity field and linear shape functions for the pressure and the viscoelastic stresses. The resulting nonlinear system of equations is solved using the Newton method. 

The problem depends on two dimensionless numbers only: the Deborah number $De=\tau\ \omega$ and the viscosity ration $\eta^*=\eta_p/\eta_s$. Measurements fix the value $\eta^*=\frac{\eta_p}{\eta_s}=1$ while the Deborah number must be estimated from the relaxation time and can be of the order of $De \approx 1.5$. We performed a continuation study by increasing the Deborah number progressively until $De=3$ above which the simulations do not converge any more. This is the consequence of the well-known high-Weissenberg number problem \cite{keunings1986high}. Nevertheless, in Fig.\ref{Fig_Marco_simu}, we show that the streamlines of the velocity field do not change qualitatively between $De=0.2$ and $De=3$ and only the magnitude of the velocity is impacted by the Deborah number. 
Therefore, small uncertainties in the estimate of the Deborah number do not change qualitatively the streamlines. Finally, simulations performed considering $h=0.5b$ confirm that changing the distance from the wall has a negligible effect on the streamlines.

\subsubsection*{Particle based simulations}

From experiments performed with pairs of microrotors, we obtained the relative velocity $\bm{v}_{exp}(r_{ji},\theta_{ji})$ between two particles as a function of $r_{ji}$ and $\theta_{ji}$ (see Fig.2 (f)). This velocity have been averaged over pairs of particles and also over time so that the back and forth motion resulting from the oscillating field is averaged to zero. The dominant contribution leading to this flow field is that generated from the first normal stress due to the particle rotation and the elastic behavior of the fluid, while the dipole-dipole interactions between the particle are negligible. 
The equation of motion of a particle then reads :

\begin{equation}\label{eq1}
\begin{gathered}
\begin{aligned}
\frac{d\bm{r}_i}{dt} &= \sum_{j\neq i} \bm{v}(r_{ji},\theta_{ji}) 
\end{aligned}
\end{gathered}
\end{equation}
with $\bm{v}$ the velocity field around a particle.
In integrating eq.\eqref{eq1}, we consider two cases for the velocity $\bm{v}$,
depending on a cut-off distance $r_c$. For $r_{ji}>r_c$, we assume that $\bm{v}=\bm{v}_{exp}/2$ where $\bm{v}_{exp}$ is an empirical function that describes the relative velocity between a pair of interacting particles measured experimentally (see Fig.4(a)). Note that to have the velocity field $\bm{v}$, one has to divide $\bm{v}_{exp}$ by 2 since $\bm{v}_{exp}$ is the result of the contribution of two particles. Thus, $\bm{v}_{exp}$ is given by:
\begin{equation}\label{eqfit}
\begin{gathered}
\begin{aligned}
v_{exp}(r_{ji},\theta_{ji}) &=  (a r_{ji} + b)\exp(-r_{ij}^2/8) + c \\
a &= (\theta_{ji}/1.22)^{2/3}(3.12+4.4) -3.12 \\
b &= -(\theta_{ji}/1.22)^{2/3}(45+24) + 45 \\
c &= -(\theta_{ji}/1.22)^{2/3}(0.3+0.4) + 0.3 \, \, \, .
\end{aligned}
\end{gathered}
\end{equation}

Since the interaction between two particles is essentially radial, we neglect the azimuthal contribution of the velocity and $\bm{v}_{exp}=v_{exp}\bm{e_r}$. 
At the end of a sequence of approach ($\theta_{ji} > \theta_l$), two particles come in close contact (side-by-side). In this situation, we observe that the pair exhibit a three-dimensional leap-frog dynamics \cite{Massana-Cid2019}, sliding on each other and ending tip-to-tip before repealing. 
To reduce the complexity of the simulation scheme, we did not consider this transitory leap-frog state, but rather take it into account as a "scattering" event. In simulations, two particles at a closer distance than $r_c$ are place at a relative position $r_{ji}=r_c$ and $\theta_{ji} = \theta_d$ in one time step:
\begin{equation}
\begin{gathered}
\begin{aligned}
v_x &= \frac{\cos(\theta_d)r_c-\Delta x}{2dt}\\
v_y &= \frac{\sin(\theta_d)r_c-\Delta y}{2dt}
\end{aligned}
\end{gathered}
\end{equation}
with $\theta_d = \pm 5^\circ$, $\Delta x = x_i - x_j$, $\Delta y = y_i - y_j$.
We perform numerical simulations of a system of $N$ point-like particles in a 2D square box of size $L$ with periodic boundary conditions. We consider a cut-off length $r_l=10 \, \rm{\mu m}$ beyond which the particles no longer interact.
The particles follow the above mentioned dynamics, and we solve the equation of motion with an Euler scheme with a time step $dt=10^{-2}$ s.

\subsubsection{Movies}
With the article are $8$ videoclips as support of the findings in the main text.
\begin{itemize}
\item \textbf{MovieS1}\textit{(.mp4)}: Oscillations of micro shakers in a PAAM solution.
The video shows the periodic back and forth movement of a dilute suspension of hematite colloidal rotors (shakers) in a polyacrylamide (PAAM) solution (0.05\% by vol). The particles are subjected to a rotating magnetic field that periodically changes sense of rotation every $\delta t=1/8$ s, see main text for Eq. (1) that gives the expression of the magnetic modulation. The applied field has amplitude $B = 5.5$ mT, frequency $f = 80$ Hz and $\delta f = 4$ Hz. The video is in real time, the scalebar is $10 \rm{\mu m}$. The video corresponds to Figure 1(a) of the article.

\item \textbf{MovieS2}\textit{(.mp4)} Absence of bands in pure water.
In this video the hematite particles are dispersed in pure water without the addition of PAAM that make the solution viscoelastic. The protocol used to apply the magnetic field and the field parameters are the same as VideoS2, however the particles do not form the bands but rather they remain dispersed in water and uniformly distributed across the observation area. Also, the field of view here is $656 \times 492 \rm{\mu m}^2$, and the video is taken in real time. 

\item \textbf{MovieS3}\textit{(.mp4)}:Band formation in a PAAM solution.
This video-clip illustrates the formation of zig-zag bands starting from a homogeneous distribution of colloidal shakers. The applied field parameters are $B = 5.5$ mT, $f = 80$ Hz and $\delta f = 4$ Hz. The video is in real time and for $t < 10$ s only the $z-$component of the rotating field is applied to induce repulsive dipolar interactions and to homogenize the particle concentration profile. After $10$ s, the full field is applied (Eq. (1) of the main text).  The field of view is $656 \times 492 \rm{\mu m}^2$, the video is in real time and corresponds to the sequence of images in Figure 1(c) of the article.

\item \textbf{MovieS4}\textit{(.mp4)} Horizontal velocity along a series of bands.
The video illustrates the time evolution of the average horizontal velocity $\langle v_y \rangle$ in a zig-zag band extracted from the positions of the colloidal shakers. The color bar on the side of the video shows the values of the velocity in $\rm{\mu m s^{-1}}$. The video illustrates the presence of vortical flows in each branch that have alternating sense of rotation. The video corresponds to Figure 1(d) of the article.

\item \textbf{MovieS5}\textit{(.mp4)} Interaction between two shakers.
The video shows the lateral speed up effect between two colloidal shakers when approach each other on the side-by-side configuration. This speed up is a manifestation of the Weissenberg effect at the microscale and emerges as strong lateral sliding when the two shakers overlap one on top of the other. The scale bar in the video is $5 \rm{\mu m}$, and the video is in real time. The video corresponds to the configurations illustrated in the schematics in Fig.2(c) of the article. 

\item \textbf{MovieS6}\textit{(.mp4)} One particle thick zig-zag band in a PAAM solution.
The video shows one band formed by one line of colloidal shakers. The particles are subjected to a rotating magnetic field with $B = 5.5$ mT, $f = 80$ Hz and $\delta f = 4$ Hz. A first particle is destabilized and dragged by the flow generated by the rest of the shakers assembled along the band (blue trajectory). The particle then switches its place with a second one which is in turn dragged by the edge flow of the band (red trajectory). The video is in real time, and it corresponds to the sequence of images in Figure 2(i) of the article.

\item \textbf{MovieS7}\textit{(.mp4)} Annihilation of two cusps in a thick band.
In this video, an ensemble of shakers forms a zig-zag band with two cusps connected by a shorter branch of rolling particles. The two cusps slowly approach by reducing the length of the connecting branch and finally merging after $30$ s of the applied magnetic driving. The used field parameters are the same as that of VideoS2 and the scale bar is $100 \rm{\mu m}$. The video is in real time, and it corresponds to the sequence of images in Figures 3(a) of the article.

\item \textbf{MovieS8}\textit{(.mp4)} Band formation from particle-based simulations.
The video shows the formation of zig-zag bands from numerical simulations. The red disks are the colloidal shakers characterized by a hydrodynamic flow field as described in the article. After a transitory, the particles organize into large scale bands displaying a constant band angle and branches with vortical flows that merges via cusps similar to the experiments.  The video corresponds to Figure 4(d) of the article.
\end{itemize}


%

\end{document}